\documentclass[]{spie}  

\usepackage{amsmath,amsfonts,amssymb}
\usepackage{graphicx}
\usepackage[colorlinks=true, allcolors=blue]{hyperref}

\usepackage{booktabs}
\usepackage[final]{microtype}

\newcommand{\0}{\hspace*{1.1ex}}

\title{Revisiting registration-based synthesis:\\A focus on unsupervised MR image synthesis}

\author[a]{Savannah~P.~Hays}
\author[a]{Lianrui Zuo}
\author[a]{Yihao Liu}
\author[a]{Anqi Feng}
\author[b]{\\Jiachen Zhuo}
\author[a]{Jerry~L.~Prince}
\author[a]{Aaron Carass}

\affil[a]{Department of Electrical and Computer Engineering, The~Johns~Hopkins~University,~Baltimore,~MD~21218}
\affil[ ]{\vspace*{-0.75em}}
\affil[b]{Department of Diagnostic Radiology and Nuclear Medicine, University~of~Maryland~School~of~Medicine, Baltimore,~MD~21201}

\authorinfo{Further author information: (Send correspondence to Savannah P. Hays)\\Savannah P. Hays: E-mail: \texttt{shays6\,@\,jhu.edu}}

\begin{document} 
\maketitle

\begin{abstract}
Deep learning~(DL) has led to significant improvements in medical image synthesis, enabling advanced image-to-image translation to generate synthetic images. 
However, DL methods face challenges such as domain shift and high demands for training data, limiting their generalizability and applicability. 
Historically, image synthesis was also carried out using deformable image registration~(DIR), a method that warps moving images of a desired modality to match the anatomy of a fixed image.
However, concerns about its speed and accuracy led to its decline in popularity.
With the recent advances of DL-based DIR, we now revisit and reinvigorate this line of research. 
In this paper, we propose a fast and accurate synthesis method based on DIR.
We use the task of synthesizing a rare magnetic resonance~(MR) sequence, white matter nulled~(WMn) T1-weighted~(T1-w) images, to demonstrate the potential of our approach. 
During training, our method learns a DIR model based on the widely available MPRAGE sequence, which is a cerebrospinal fluid nulled~(CSFn) T1-w inversion recovery gradient echo pulse sequence.
During testing, the trained DIR model is first applied to estimate the deformation between moving and fixed CSFn images. 
Subsequently, this estimated deformation is applied to align the paired WMn counterpart of the moving CSFn image, yielding a synthetic WMn image for the fixed CSFn image.
Our experiments demonstrate promising results for unsupervised image synthesis using DIR.
These findings highlight the potential of our technique in contexts where supervised synthesis methods are constrained by limited training data.
\end{abstract}


\section{INTRODUCTION}
\label{sec:intro}  
Image synthesis is typically achieved using image-to-image translation~(I2I), which takes a source domain image as input and learns a regression function $f(\cdot)$ to generate a synthetic image in a target domain. 
The underlying anatomical information of the input image is preserved during I2I. 
Depending on the training data, these I2I-based synthesis methods can be categorized into supervised and unsupervised approaches~\cite{yang2020tmi, zuo2021unsupervised, zuo2022haca3}.
Supervised I2I-based methods rely on paired data, where the anatomy of the source and target training images match. 
For example, a supervised I2I-based model can learn to generate synthetic T1-weighted~(T1-w) magnetic resonance~(MR) images from paired T2-weighted~(T2-w) images~\cite{zuo2020synthesizing}.
On the other hand, unsupervised methods, such as CycleGAN~\cite{zhu2017unpaired}, do not require paired source and target images for training. 
Despite the success of I2I-based image synthesis, both supervised and unsupervised methods face several challenges. 
Supervised I2I-based methods are constrained by their reliance on ample paired data for training, limiting their applicability in settings with scarce paired scans. 
Unsupervised I2I-based methods, on the other hand, have difficulty maintaining the geometric fidelity of anatomical structures during synthesis~\cite{yang2020tmi, zuo2021unsupervised, zuo2022haca3}.

Image synthesis has also been accomplished in the past using deformable image registration~(DIR)~\cite{miller1993mathematical, hofmann2008jnm, hofmann2011jnm, burgos2014tmi, lee2017spie}.
These methods achieve image synthesis by warping moving images of a desired domain to match the anatomy of a fixed image. 
As a result, the synthetic image retains the domain information of the moving image and the anatomical information of the fixed image.
This approach allows us to synthesize images in an unsupervised fashion~(i.e., without matched anatomy across domains).
The same approach has also been used for atlas-based segmentation~\cite{hofmann2011jnm}.
However, the use of DIR-based synthesis has dwindled in recent years due to concerns regarding their speed and accuracy compared with I2I-based synthesis.

There have been recent advances in deep learning~(DL)-based DIR with respect to both speed and accuracy; see the review article of Chen \textit{et al.}~\cite{chen2023arxiv} for more details.
Building upon these improvements, we revisit and reinvigorate DIR-based methods for image synthesis. 
In particular, we propose a fast and accurate DIR-based synthesis method and demonstrate the potential of this approach by synthesizing white matter nulled~(WMn) T1-w images, an MR contrast that is increasingly recognized for deep gray matter regions, but not commonly acquired.  
The scarcity of this sequence implies that there is limited training data, which creates a problem for both supervised and unsupervised WMn T1-w image synthesis.
Our proposed method addresses this challenge by first training a DIR model on widely available cerebrospinal fluid nulled~(CSFn) T1-w magnetization-prepared gradient echo~(MPRAGE) images.
In the testing phase, we first use the trained DIR model to estimate the deformation between a moving CSFn image and a fixed CSFn image. 
Subsequently, we apply this deformation to warp the paired WMn image counterpart of the moving CSFn image, aligning it to the fixed CSFn image. 
This process enables the synthesis of a new WMn image with the anatomy of the fixed CSFn image.
Our results demonstrate that our method outperforms other DIR-based synthesis methods while maintaining enhanced generalizability with respect to other I2I-based synthesis methods.
The contributions of the work are the following. 
First, we provide a comprehensive review and comparison of I2I-based and DIR-based methods for WMn image synthesis. 
Secondly, we present a novel DIR-based method for WMn image synthesis with improved speed and accuracy. 
Lastly, we provide insights about the unique benefits of DIR-based synthesis, especially in handling limited data.

\begin{figure}[!tb]
\centering
\includegraphics[width=0.99\textwidth]{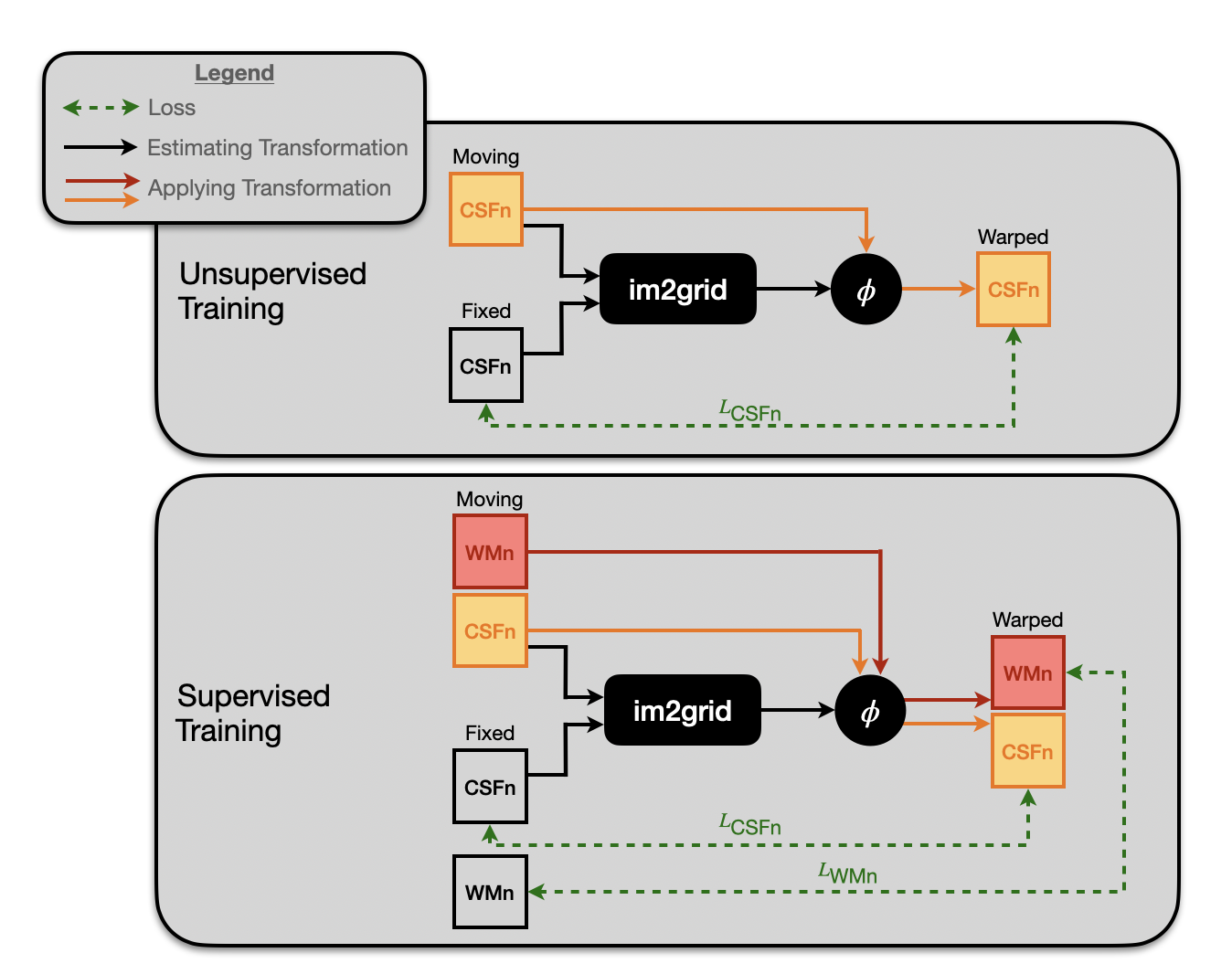} 
\caption{Framework of training our proposed method. During unsupervised training (top), the MSE loss is calculated between the fixed and warped CSFn images. During supervised training (bottom), the MSE loss is calculated between the fixed and warped images for both CSFn and WMn acquisitions. Green dashed lines denote the loss computations.}
\label{fig:registrationtraining}
\end{figure}

\begin{figure}[!tb]
\centering
\includegraphics[width=0.99\textwidth]{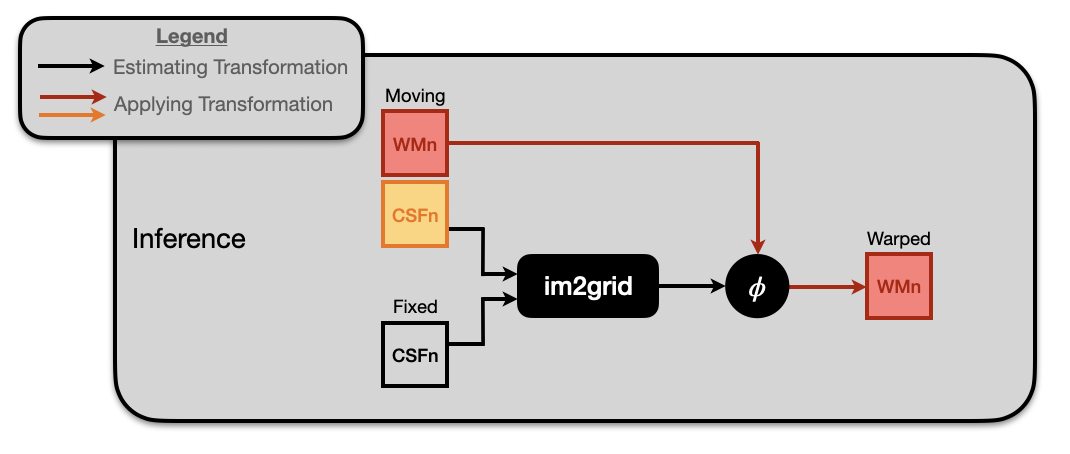} 
\caption{Framework of testing our proposed method. The transformation between $m_\text{CSFn}$ and $f_\text{CSFn}$ is first estimated. The estimated transformation is then applied to the moving WMn image, $m_\text{WMn}$, resulting in a synthetic WMn image for $f_\text{CSFn}$.}
\label{fig:testing}
\end{figure}

\section{METHODS}
\subsection{WMn image synthesis based on I2I}
\label{sec:methodsI2I}
Our data consists of two MR tissue contrasts---CSFn MPRAGE and WMn MPRAGE---acquired from 47~subjects at the University of Maryland School of Medicine.
The CSFn and WMn images were acquired contemporaneously with identical imaging sequence parameters except the inversion time~(TI) which was 1,400ms and 400ms, respectively.
The MR images were preprocessed following Tohidi \textit{et al}~\cite{tohidi2023synthesis}.
For comparison, we trained both a supervised I2I method based on a 3D U-Net~\cite{ronneberger2015u} and an unsupervised I2I method based on a  CycleGAN~\cite{zhu2017unpaired} with the CSFn image as input and the WMn image as the desired output.
For both methods, 30 subjects were used for training, 7 for validation, and 10 for testing.

\subsection{DIR-based synthesis}
We adopted the recently developed DL-based DIR algorithm \textit{im2grid}~\cite{liu2022coordinate} for learning the registration as shown in Fig.~\ref{fig:registrationtraining}.
The inputs to the \textit{im2grid} network are two CSFn images: the fixed, $f_\text{CSFn}$, and moving, $m_\text{CSFn}$, image.
The network predicts the transformation that aligns $m_\text{CSFn}$ to $f_\text{CSFn}$.
During training, \textit{im2grid} incorporates a differentiable grid sampler~\cite{jaderberg2015spatial} to produce a warped CSFn image, $w_\text{CSFn}$, from $m_\text{CSFn}$.
This allows the network to be trained in an unsupervised manner using the dissimilarity between $w_\text{CSFn}$ and $f_\text{CSFn}$.
We can train this network in a supervised manner using an additional loss on the paired fixed MWn, $f_\text{WMn}$, image and warped WMn, $w_\text{WMn}$, image.
At test time, a transformation is predicted by 
aligning $m_\text{CSFn}$ and $f_\text{CSFn}$ as shown in Fig.~\ref{fig:testing}.
The predicted transformation is then applied to the moving WMn image, $m_\text{WMn}$, that paired with $m_\text{CSFn}$, yielding a synthetic WMn image, $w_\text{WMn}$, for $f_\text{CSFn}$.

An advantage of using a DL algorithm for DIR is the flexibility in incorporating various loss functions.
Unlike traditional registration algorithms such as SyN~\cite{avants2008symmetric}, deep learning allows easy integration of additional losses, such as the label-wise Dice loss. 
DL-based algorithms for DIR can also achieve supervised training when the paired WMn images are available; supervision for the synthesis can be incorporated using an extra dissimilarity loss between $w_\text{WMn}$ and $f_\text{WMn}$.
A visualization of training our method is shown in Fig.~\ref{fig:registrationtraining}.
In our experimental setup, we adopted the same training, validation, and testing split as in Sec.~\ref{sec:methodsI2I}.
The supervised version of our proposed method trains \textit{im2grid} with MSE losses between the CSFn images, $f_\text{CSFn}$ and $w_\text{CSFn}$, and the WMn images, $f_\text{WMn}$ and $w_\text{WMn}$, and a Dice loss from 127 segmentation labels on the CSFn images,  $f_\text{CSFn}$ and $w_\text{CSFn}$; the 127 labels come from application of SLANT~\cite{huo2016spie, huo20193d} to the CSFn image.
The unsupervised version of our proposed method trains \textit{im2grid} using MSE loss on only the CSFn images, $f_\text{CSFn}$ and $w_\text{CSFn}$, as well as the Dice loss from the 127 SLANT segmentation labels on the CSFn images, $f_\text{CSFn}$ and $w_\text{CSFn}$.
The incorporation of Dice loss further enhances the structural similarity between the fixed and warped images.

\begin{figure}[!tb]
        \centering
        \includegraphics[width=0.66\textwidth]{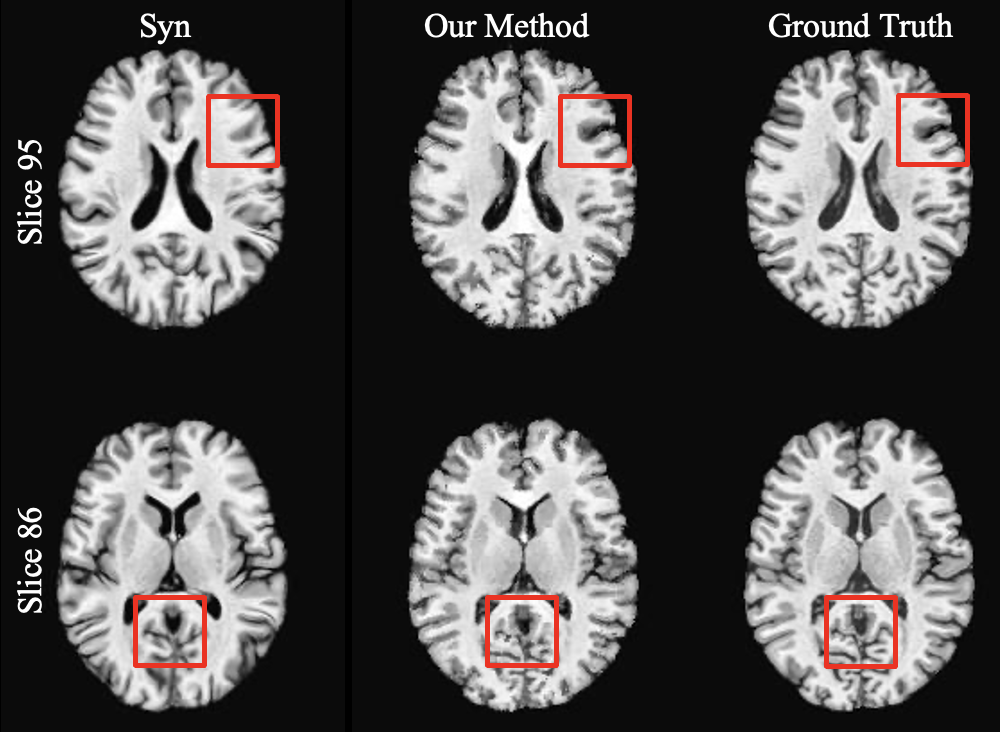} 
        \caption{Visualization of the warped CSFn images using SyN and our method with one moving image. The red boxes highlight regions where SyN does not perform well.}
    \label{fig:warped_mprages}
\end{figure}

\section{EXPERIMENTS AND RESULTS}
\subsection{Validating CSFn registration}
In DIR-based synthesis, a deformation is learned by warping CSFn images $m_\text{CSFn}$ to $f_\text{CSFn}$ before it is applied to $m_\text{WMn}$.
Before evaluating the performance of synthesizing WMn images using DIR, we first validate the learned deformation based on CSFn registration.
As a comparison to our DIR-based method, we used SyN~\cite{avants2008symmetric} implemented in the ANTs package~\cite{avants2009ants}.
The same ten testing subjects were used for evaluation of the I2I-based synthesis and DIR-based synthesis methods.
In Fig.~\ref{fig:warped_mprages}, we show the warped CSFn images of two different slices of the same moving subject for both SyN and our unsupervised DIR-based method.
The red boxes highlight major differences, where SyN did not perform well in comparison to our method.
Evaluating how well our model can perform in this first step of warping $m_\text{CSFn}$ to $f_\text{CSFn}$ is associated with the accuracy of the warped WMn image, $w_\text{WMn}$.
For SyN, the structural similarity index measure~(SSIM) and the peak signal-to-noise ratio (PSNR) of the warped CSFn image was $0.8287\pm0.007$ and $ 23.52\pm1.535$, respectively.
For our method, the SSIM and the PSNR of the warped CSFn image was $0.9758\pm0.003$ and $35.85\pm0.5283$, respectively.
These metrics were calculated excluding the background pixels and using one moving image.

\subsection{WMn image synthesis}
\begin{figure}[!tb]
    \centering
    \includegraphics[width=0.9\textwidth]{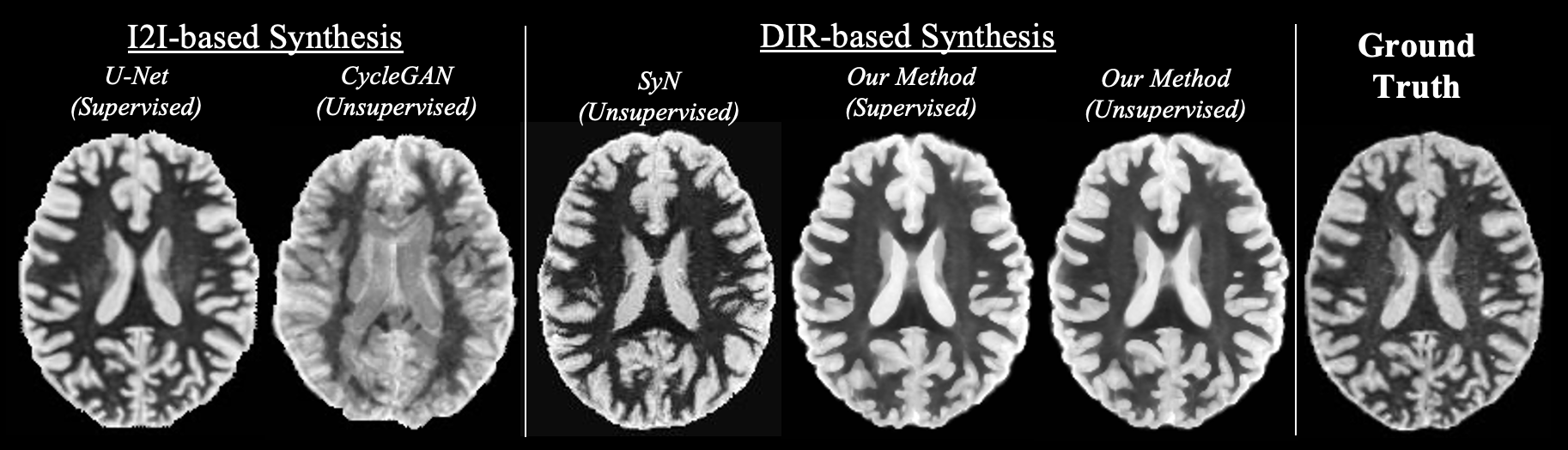}
    \caption{Visualization of the synthetic WMn images using five different approaches.}
    \label{fig:warped_fgatirs}
\end{figure}

In Fig.~\ref{fig:warped_fgatirs}, we visually compare the WMn synthesis results from the I2I-based synthesis and DIR-based synthesis methods.
In Table~\ref{tab:fgatir_metrics}, we report the SSIM and PSNR of the five synthesis methods including our two DIR-based methods: supervised and unsupervised.
Due to the fast inference time of our model, we were able to synthesize multiple WMn images for one fixed subject within a minute.
The performance of our DIR-based method was boosted when using multiple moving CSFn images from different subjects to synthesize a warped WMn image for a single fixed CSFn image.
To synthesize a single warped WMn image, we took the pixel-wise mean of the multiple warped WMn images, which yielded a better result than the median.
Due to our limited dataset, the maximum number of moving images we could explore was nine, which yielded the best result for both our models~(supervised and unsupervised).
In unsupervised training the DIR-model learns the deformation field between two CSFn images and the WMn images are not used.
Therefore, the WMn images of our training subjects can be used during testing as moving images since we are measuring the dissimilarity between two WMn images and not the dissimilarity between the CSFn images.
We take advantage of this usage later in Section~\ref{sec:movingimages} where we show results with greater than nine moving images.
The supervised U-Net outperforms our supervised DIR-based method, but our unsupervised DIR-based method outperformed CycleGAN.
The CycleGAN model did not converge well for this task.
We believe this is due to the unique image contrast of the WMn images, the limited dataset, and training on full axial slices (excluding slices with limited brain tissue).

\begin{table}[!tb]
    \caption{Numerical comparison of I2I-based and DIR-based methods in WMn image synthesis. ``Proposed~(S)'' is the supervised version of the proposed method, where \textit{im2grid} was trained on both CSFn and WMn images. ``Proposed~(U)'' stands for the unsupervised version of the proposed method, where \textit{im2grid} was only trained on CSFn images.}
    \label{tab:fgatir_metrics}
    \centering
    \begin{tabular}{l cc cc cc cc cc}
    \\[-0.7em]
        \toprule
        & \multicolumn{4}{c}{\bfseries{I2I-based}} & \multicolumn{6}{c}{\bfseries{DIR-based}}\\
        \cmidrule(lr){2-5}
        \cmidrule(lr){6-11}
        && \bfseries{U-Net} && \bfseries{CycleGAN} && \textbf{SyN} && \textbf{Proposed~(S)} && \textbf{Proposed~(U)}\\
        \cmidrule(lr){2-5}
        \cmidrule(lr){6-11}
        \bfseries{SSIM} && $\0{}0.83\pm0.857$ && $\0{}0.64\pm0.027$ && $\0{}0.70\pm0.071$ && $\0{}0.79\pm0.059$ && $\0{}0.81\pm0.017$\\
        \cmidrule(lr){2-5}
        \cmidrule(lr){6-11}
        \bfseries{PSNR} && $25.13\pm1.213$ && $16.72\pm0.216$ && $20.19\pm2.020$ && $21.24\pm1.627$ && $21.99\pm1.187$\\
        \bottomrule
    \end{tabular}
\end{table}

\subsection{Multi-contrast MR image synthesis}
\begin{figure}[!tb]
    \centering
    \includegraphics[width=0.9\textwidth]{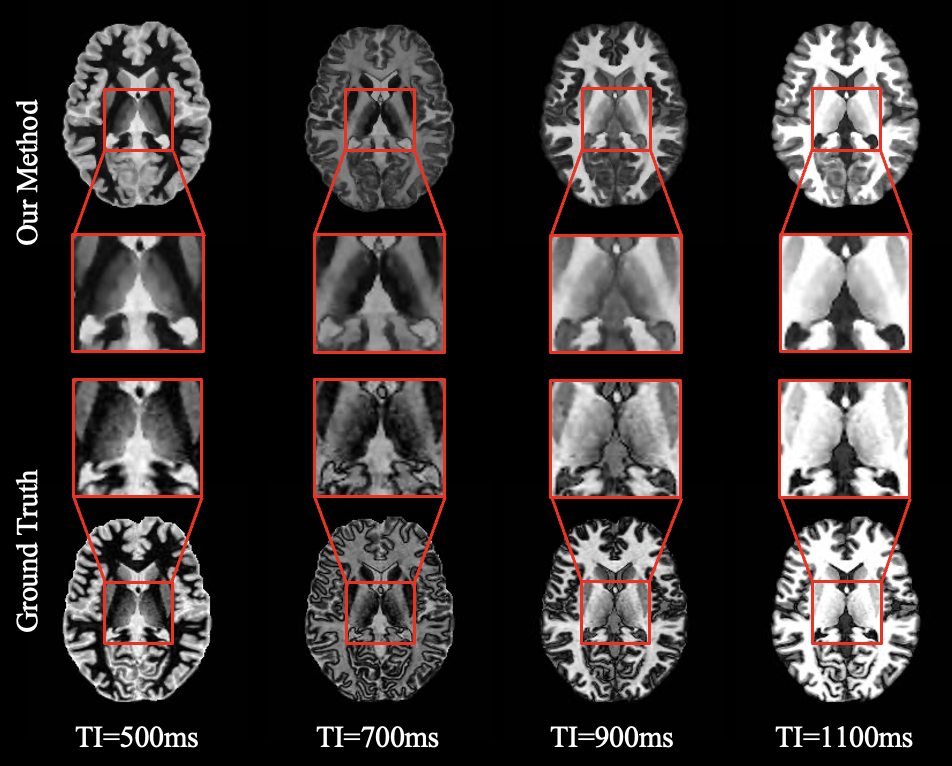}
    \caption{Synthetic multi-TI images using our unsupervised DIR-based model without retraining. We show a specific slice that highlights the preservation of the thalamus structure. Nine moving images were used in this experiment to generate the warped images.}
    \label{fig:warped_multiTI}
\end{figure}

A highlight of our unsupervised DIR-based method is the predicted transformation by the model can be applied to additional image contrasts allowing us to synthesize multiple image contrasts with only one model, which was trained using only CSFn images.
In Fig.~\ref{fig:warped_multiTI} we show results for synthesizing multi-TI images using the same unsupervised DIR-based method without retraining.
The moving multi-TI images were calculated based on our paired CSFn and WMn images which has been previously shown to be feasible in Tohidi \textit{et al}~\cite{tohidi2023synthesis}.
Different TI images can be useful to highlight different anatomical structures, we have included four different TI values in Fig.~\ref{fig:warped_multiTI} to demonstrate this point.
Our method yields synthetic images with strong visibility of the thalamus similar to the corresponding ground truth images.

\subsection{Investigating supervised and unsupervised DIR-based synthesis}
\begin{figure}[!tb]
    \centering
    \includegraphics[width=0.9\textwidth]{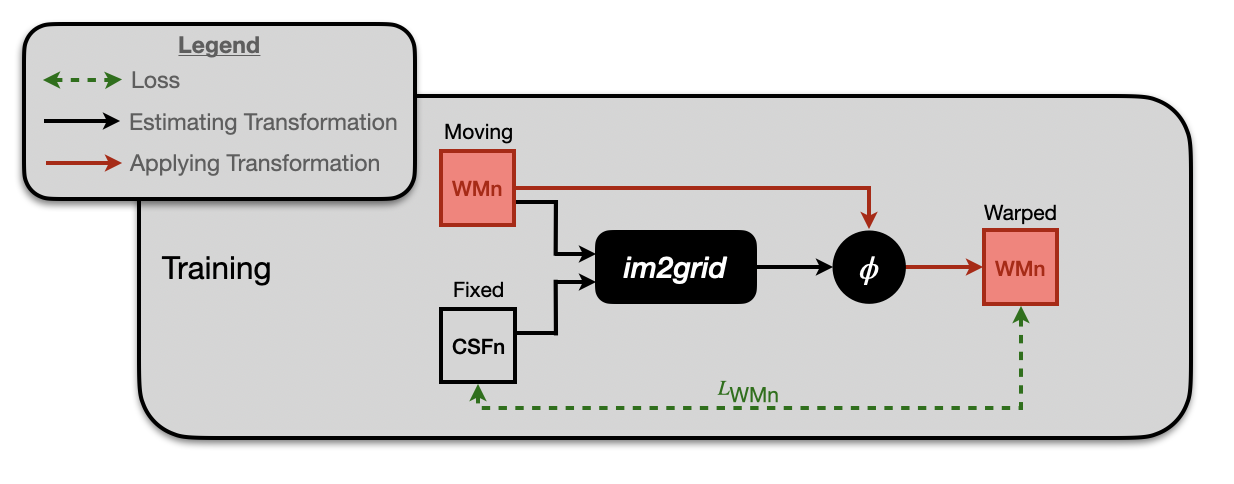}
    \caption{Framework for training our model for only WMn image registration.}
    \label{fig:fg_diagram}
\end{figure}

\begin{figure}[!tb]
    \centering
    \includegraphics[width=0.66\textwidth]{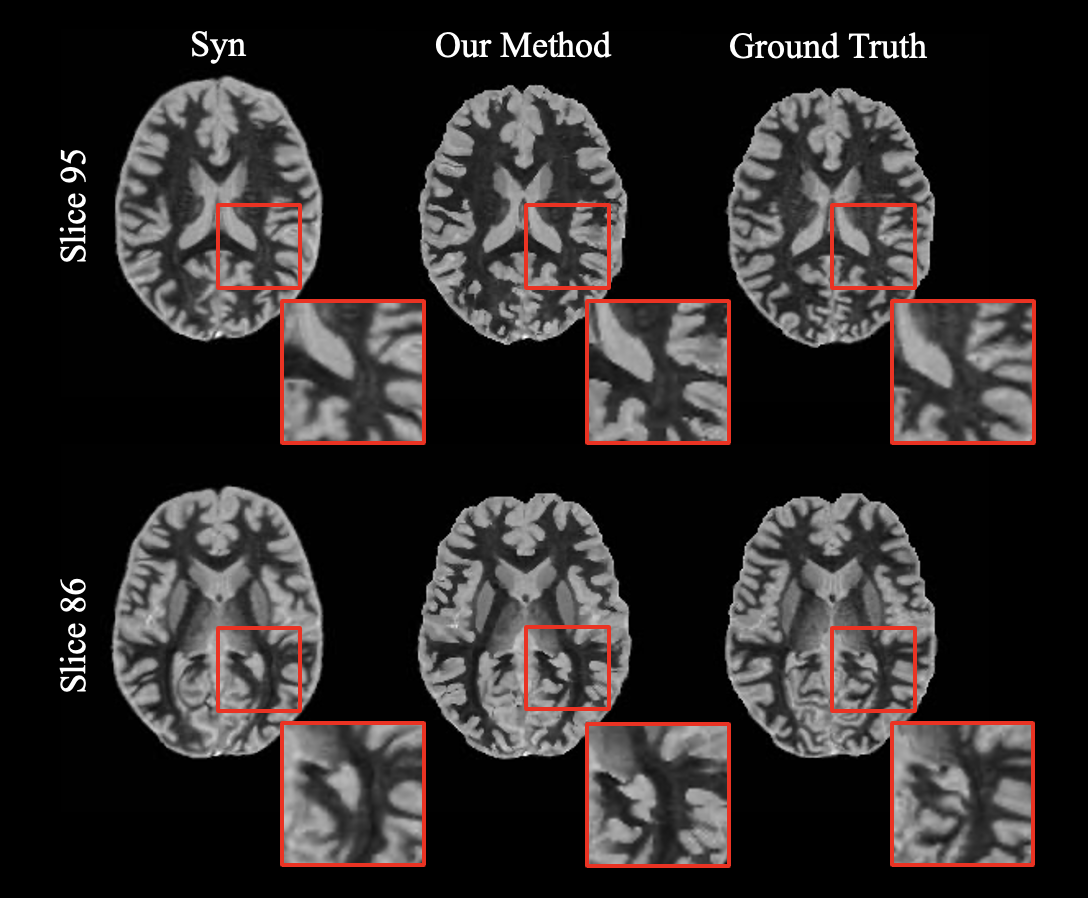}
    \caption{Visualization of the warped WMn images using SyN and our method with one moving image.}
    \label{fig:reg_fg}
\end{figure}

For our supervised DIR-based method, we found that it does not outperform the unsupervised DIR-based method.
This is a surprising result since one would expect adding an additional loss based on WMn would improve DIR-based synthesis.
To understand a possible reason for this, we trained a model using the WMn images as input to the \textit{im2grid} network instead of the CSFn images as shown in Fig.~\ref{fig:fg_diagram}.
We identify that both our method and ANTs struggle to accurately warp WMn images, as shown in Fig.~\ref{fig:reg_fg}.
Our method achieved a PSNR of $22.71\pm1.3321$ and SSIM of $0.64\pm0.002$.
ANTs achieved a higher PSNR of $23.09\pm0.6135$ but a lower SSIM of $0.6129\pm0.004$.
This is very different from the results of the CSFn images shown in Fig.~\ref{fig:warped_mprages}.
We believe that the unique contrast of the WMn images plays a role in the degradation of performance for both our method and SyN.

\subsection{Impact of the moving images} 
\label{sec:movingimages}
\begin{figure}[!tb]
    \centering
    \includegraphics[width=0.9\textwidth]{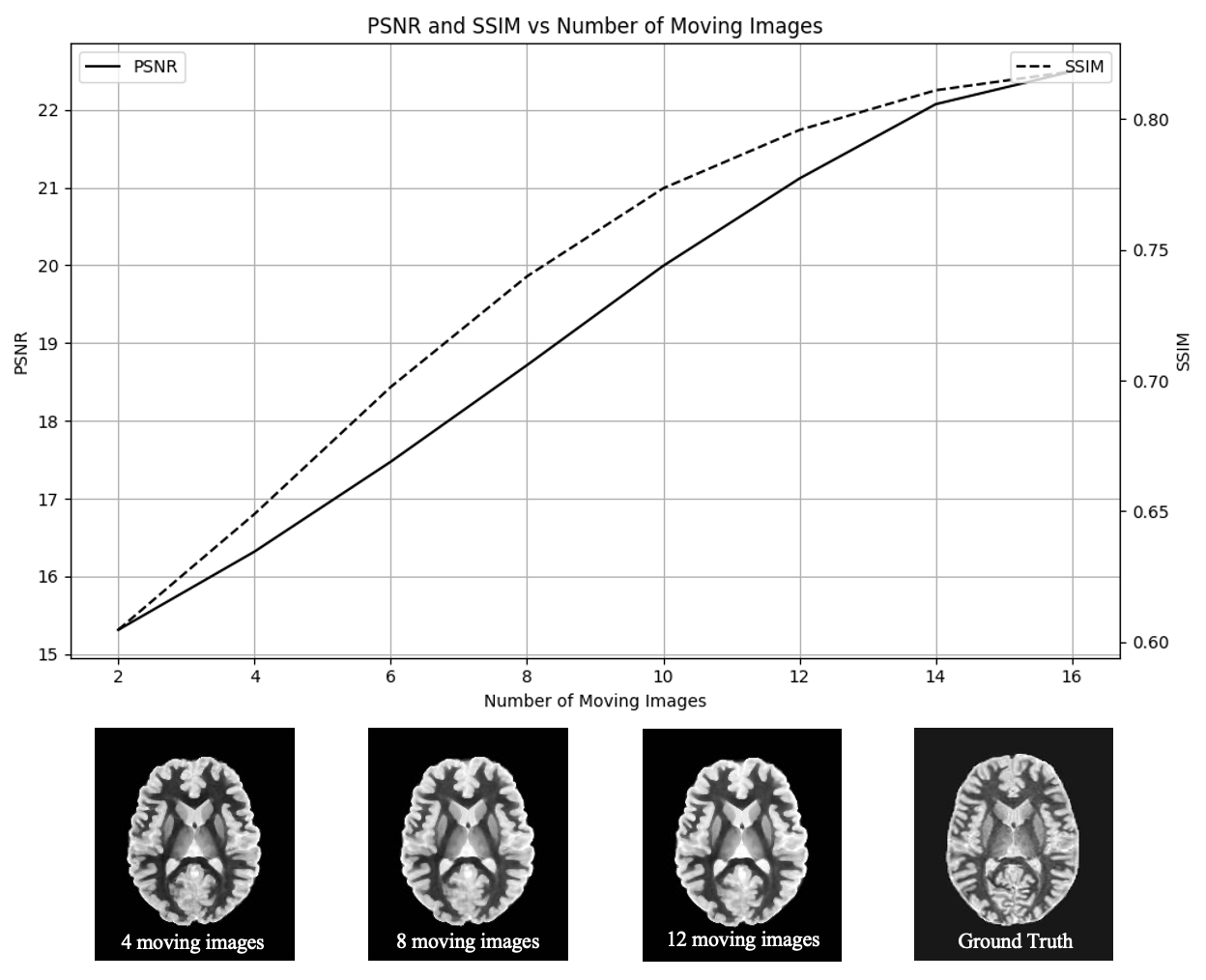}
    \caption{The impact of the number of moving images used in the mean warped image on the PSNR and SSIM metrics. The top row show a plot of the PSNR and SSIM performance of our method as you increase the number of moving images. The bottom row show example output images from our method and the corresponding ground truth image.}
    \label{fig:moving_images}
\end{figure}

We previously mentioned that our unsupervised DIR-based method does not require the synthesis target, WMn images, during training.
During testing, we can use our available WMn images as moving images to synthesize a warped WMn image.
We explored the impact of using multiple moving images and taking the pixel-wise mean to generate the synthetic warped WMn image.
The PSNR and SSIM of different quantities of moving images are shown in Fig.~\ref{fig:moving_images}.
Interestingly, as the number of moving images increases the PSNR and SSIM increases.
From this graph, we can infer the expected performance of our unsupervised DIR-based model based on the availability of the synthesis target contrast images.
For example, if we have 8 available MR images of the synthesis target contrast then we would expect to achieve a PSNR around $18.75$ and SSIM around $0.74$ when using our unsupervised DIR-based method.

\section{Discussion and Conclusion}
In this work, we revisited DIR-based synthesis, which has taken a backseat in recent years due to the emergence of direct DL-based synthesis techniques.
While supervised I2I-based methods have shown superior performance, our results demonstrate the potential of unsupervised DIR-based synthesis and its improved performance compared to other registration methods such as SyN.
The advantage of our method lies in its independence from the synthesis target during training.
This characteristic makes our method particularly useful when working with limited datasets.
We are able train our model on an abundantly acquired MR image contrast then take full advantage of our rarely acquired, limited image contrast during testing time.
Another notable advantage of our method is its run time efficiency during testing, as compared to SyN.
While SyN takes over two hours to warp a single image to a fixed image, our method only requires a few seconds.
This swift processing time enables us to potentially warp multiple atlases to the fixed image, facilitating the synthesis of more accurate images.
A unique advantage to our unsupervised model is it does not need to be retrained to synthesize other MR modalities that have paired CSFn images.
I2I-based synthesis models both supervised and unsupervised need to be retrained given the desired synthesis target.
Our results show a promising future of DIR-based synthesis in the era of DL that could lead to overcoming issues of generalizability and limited datasets.

In training our DIR-based model, the first step involves learning the deformation field between two CSFn images.
This is a major limitation of this method.
The synthetic warped WMn image will not have better accuracy than the warped CSFn image.
There are many parameters that need to be explored in this DIR-based synthesis method to achieve optimal performance.
When synthesizing a warped image using multiple moving images in this work, we used a pixel-wise mean.
There are more complex ways to approach this combination that could produce a more accurate synthetic image.
One potential solution is to compare the SLANT segmentation of the moving images with the SLANT segmentation of the fixed image and use a weighted pixel-wise mean of the moving images where favor is given to the images with a more similar segmentation to the fixed image.
Therefore the moving images that have a more similar anatomy to the fixed image will be favored.
Previous methods have been developed that take advantage of local features to weight images differently such as joint label fusion \cite{wang2013ieee} and non-local statistical label fusion \cite{asman2013mia} for multi-atlas segmentation.
Our DIR-based approach provides an avenue for exploration to apply these fusion methods in a synthesis task.
Since our model is registration-based, it may miss pathology changes that are only shown in the WMn images.
Overcoming pathological differences is a problem present in both DIR-based and I2I-based synthesis.

When training our DIR-based model, we equally weighted the loss functions.
We explored with DSC, MSE, and normalized cross correlation (NCC) losses and found MSE to give a slightly better result than NCC for our specific task and dataset.
The optimal combination of the losses and their respective weightings is another avenue for exploration, but it is important to remember that the incorporation of the DSC loss is one explanation for why our DIR-based model heavily outperforms SyN.


\acknowledgments 
 
This work was supported in part by the NIH through the National Institute of Neurological Disorders and Stroke~(NINDS) grant R01-NS105503 (PI: R.P. Gullapalli). The data was acquired in line with the principles of the Declaration of Helsinki. Approval was granted by an IRB Committee of the University of Maryland, School of Medicine.

\bibliography{cas-refs} 
\bibliographystyle{spiebib} 

\end{document}